# Towards a Semantic Preservation System

**A Technical Note prepared as part of the National Archives and Records Administration (NARA) funded "Innovative Systems and Software: Applications to NARA Research Problems" project.**

Robert E. McGrath, Jason Kastner, Alejandro Rodriguez, Jim Myers
**Cyberenvironments and Technologies Directorate**
**National Center for Supercomputing Applications**
**University of Illinois, Urbana-Champaign**

# Towards a Semantic Preservation System

**Towards a Semantic Preservation System**

# Abstract

This document discusses the status of the Defuddle parser and recent work conducted as part of the "Innovative Systems and Software: Applications to NARA Research Problems" project.

Preserving access to file content requires preserving not just bits but also meaningful logical structures. The ongoing development of the Data Format Description Language (DFDL) is a completely general standard that addresses this need. The Defuddle parser is a generic parser that can use DFDL-style format descriptions to extract logical structures from ASCII or binary files written in those formats. DFDL and Defuddle provide a preservation capability that has minimal format-specific software and cleanly separates issues related to bits, formats, and logical content. Such a system has the potential to greatly reduce overall system development and maintenance costs as well as the per-file-format costs for long term preservation.

This project is investigating extending this model to extract descriptions of the structure and relations in the data into standard semantic web languages (the Resource Description Framework (RDF) (http://www.w3.org/TR/rdf-concepts/) and the Web Ontology Language (OWL) (http://www.w3.org/TR/owl-features/). Our approach is a two-step process, the standard DFDL processing within Defuddle to generate XML, and a second phase to extract semantic descriptions from the XML using the Gleaning Resource Descriptions from Dialects of Languages (GRDDL) standard (http://www.w3.org/TR/grddl/) as a standard mechanism for declaring these transformations.

This document provides a brief overview of DFDL's scope and reviews the basic design of the Defuddle parser. We have recently done work that includes updates to Defuddle, investigation of its performance characteristics, extensions to support semantic file descriptions (i.e. using RDF and community ontologies), and support for integration with SHAMAN as an iRODS microservice.

# Towards a Semantic Preservation System

**Towards a Semantic Preservation System**

# Contents

# Towards a Semantic Preservation System

# 1. Introduction

Preservation can be thought of as communication with the future. The records we preserve today need to be accessible and displayable by future technology. Beyond maintaining the accessibility of the raw bits of the digital data, preservation requires maintaining an ability to interpret the data as meaningful structures and relationships and displaying accurate visual representations of them to our descendants. Preserving access to file content in terms of meaningful logical structures is a primary motivation for the ongoing development of the Data Format Description Language (DFDL) and Defuddle – a generic parser that can use DFDL-style format descriptions to extract logical structures from ASCII or binary files written in those formats. Combined with capabilities for long-term preservation of raw bits and a viewer for logical structures as is being done within the European Union's Sustaining Heritage Access through Multivalent Archiving project (SHAMAN [14]), DFDL and Defuddle provide an end-to-end preservation capability that has minimal format-specific software and cleanly separates issues related to bits, formats, and logical content. Such a system has the potential to greatly reduce overall system development and maintenance costs as well as the per-file-format costs for long term preservation.

In addition to direct use in preservation, DFDL/Defuddle can provide value in the overall curation and preservation process and support data sharing and reuse in an e-Science context. For example, DFDL can be used to describe the logical structures used to recognize file MIME types, enabling Defuddle to serve as a MIME type identifier. It can also be used to describe the logical structure of descriptive metadata embedded in those files, thus enabling Defuddle to operate as a metadata extractor. Thus these operations can also be performed using format-independent software and declarative per-format descriptors to minimize long-term costs.

Understanding DFDL/Defuddle in this context, that it can be used to map formats to multiple, use-specific logical models, provides a sense of its broader potential in eScience. DFDL/Defuddle could be used to define common logical models present in multiple file formats to allow integration of syntactically heterogeneous data or harvest the common subset of data available across multiple file types. (With the addition of an 'inverse parser' to go from logical models to physical formats, it could serve as a file translator.) Thus DFDL/Defuddle can serve as a community-scale coordination mechanism to support metadata-based discovery systems and/or the creation of reference collections without requiring full format standardization, and consequent standardization, of the software that produces, analyzes, and visualizes the data across the broad community(ies) involved.

As detailed later in this report, the current version of DFDL is defined as a set of allowed annotations of an XML schema, the latter of which captures the 'logical model'. While the XML Schema language is well suited for describing the layout of data (the "syntax"), interoperability and robust archiving require semantic mark up as well. Specifically, XML is hierarchical where logical models may have more general relationships between models. Further, XML identifiers are local to the current document and cannot easily support, for example, association of logical model elements with external resources (e.g., annotation).  This project seeks to extend this model to extract descriptions of the structure and relations in the data into standard semantic web languages (the Resource Description Framework (RDF) and the Web Ontology Language (OWL). In addition to providing a richer framework for defining logical models and supporting global identifiers, an RDF/OWL analog of DFDL would be more amenable to logical inference and the use of rules to automate further enrichment of logical models and their associations with other resources.

This document provides a brief overview of DFDL's scope and reviews the basic design of the Defuddle parser. It also discusses the status of the Defuddle parser and our efforts as part of the "Innovative Systems and Software: Applications to NARA Research Problems" project to update it, understand its performance characteristics, extend it to support semantic file descriptions (i.e. using RDF and community ontologies) and integrate it with SHAMAN as an iRODS microservice. Section 2 introduces DFDL and Defuddle. Section 3 describes accomplishments and ongoing work related to core Defuddle, its semantic extensions, and its integration into the SHAMAN architecture. Section 4 gives a detailed example of Defuddle, including semantic extensions. Section 5 describes the performance characteristics of Defuddle and provides performance numbers and scaling behavior for some simple cases. Section 6 closes with implications and future work.

# 2. Background

The Data Format Description Language (DFDL) is a draft standard specification from the Open Grid Forum (http://forge.gridforum.org/projects/dfdl-wg).[1] DFDL proposes to describe existing data formats, both binary and text, in a manner that makes the data accessible through generic mechanisms. The DFDL specification is based on XML Schema, a standard that defines the structure and semantics of XML document formats. In analogy with an XML parser, which can use an XML schema to interpret XML input in terms of the logical model in the schema, a DFDL parser can interpret an input that is a sequence of bytes (ASCII or binary, not necessarily XML) in terms of schema. In both cases, the output is an XML Information Model (for DFDL, *as if* the input was XML).

XML Schema allows annotation of schemas, a feature primarily used for the benefit of human readers which is also usable by applications. DFDL uses this mechanism to add information regarding the application of the formal structure of an XML schema to arbitrary data file formats. DFDL annotations specify low-level format issues such as whether the source format is ASCII or binary and, for binary, whether "big-endian" or "little-endian" encodings have been used. They also specify higher level associations between the raw bytes on the disk and the logical data model specified by the XML schema, including which bytes are associated with which XML elements and how to interpret bytes in a particular format as control structures. For example, as described by Talbott et.al. [15], the DFDL language supports:

- Conditional logic (if, choice, any)
- Basic math operations (+,-,*,/)
- Looping
- Pattern matching for text/binary delimiters
- Reference values within schema (for sequence length, delimiters, etc.), and
- Layering (hidden elements that can be referenced, but do not display in output)

[1] The DFDL specification is still under development and has not released an official draft. The discussion in this section is based on an early draft of the specification which includes features that have since been removed from consideration as part of a version 1.0 specification but are expected to be re-introduced in later iterations,

Additional capabilities of DFDL include the ability to reference external algorithms as part of the overall mapping (e.g. compression and/or encryption algorithms), the ability to consider multiple files to be part of a single logical data format (e.g. as is the case with the “shape” (*.shp) format for geospatial information), and the ability to annotate XML ComplexTypes to support re-use of mappings for common logical structures shared by multiple formats (e.g. for a comma-separated value (CSV) table embedded in a multicomponent file). DFDL also inherits XML schema’s capabilities to add additional information (e.g. static attributes that might specify the physical units associated with a value), and place constraints on values and numbers of elements, which, respectively, allow association of additional knowledge about the file format to be associated with instance data without bloating source data files, and for source files to be checked for conformance with the format specification.

The open source Defuddle parser [15] was originally developed within the DOE-funded Scientific Annotation Middleware (SAM) project (www.scidac.org/SAM/) as both a proof of concept of the DFDL specification and a mechanism for testing concepts which can feed back into the specification process. A specific aim of Defuddle is to demonstrate that an efficient, generic DFDL parser can be built and that such a parser can effectively address real-world examples.

Defuddle extends the concepts embedded in the Java XML Binding (JAXB) Specification), which specifies a means of using an XML schema to automatically generate Java classes to represent data conforming to that schema and to parse a stream of XML data (e.g. from an XML file) to instantiate a set of Java objects to represent that data in memory. Defuddle uses DFDL annotations on such a schema as a way to extend the generated classes and parsing logic to support an ASCII or binary data stream rather than XML. Specifically, Defuddle extends the open source Apache JaxMe (http://ws.apache.org/jaxme/) implementation of JAXB and follows the same 2-step model (Figure 1) of parsing the schema (XML Schema or DFDL format description for JaxMe and Defuddle respectively) to generate Java source code for the classes required to parse and represent data in the relevant format followed by a step in which those classes are compiled and run on instances of that format. In both cases, the result is an in-memory representation of the data that is logically equivalent to the model expressed in the schema and which can be manipulated directly within an application or exported into a new XML file. (For JaxMe, this is a round trip from XML to a canonically equivalent new XML document. For Defuddle, it is a translation from ASCII/binary to XML.) Thus data can be manipulated according to its logical structure (specified by the XML schema), regardless of how

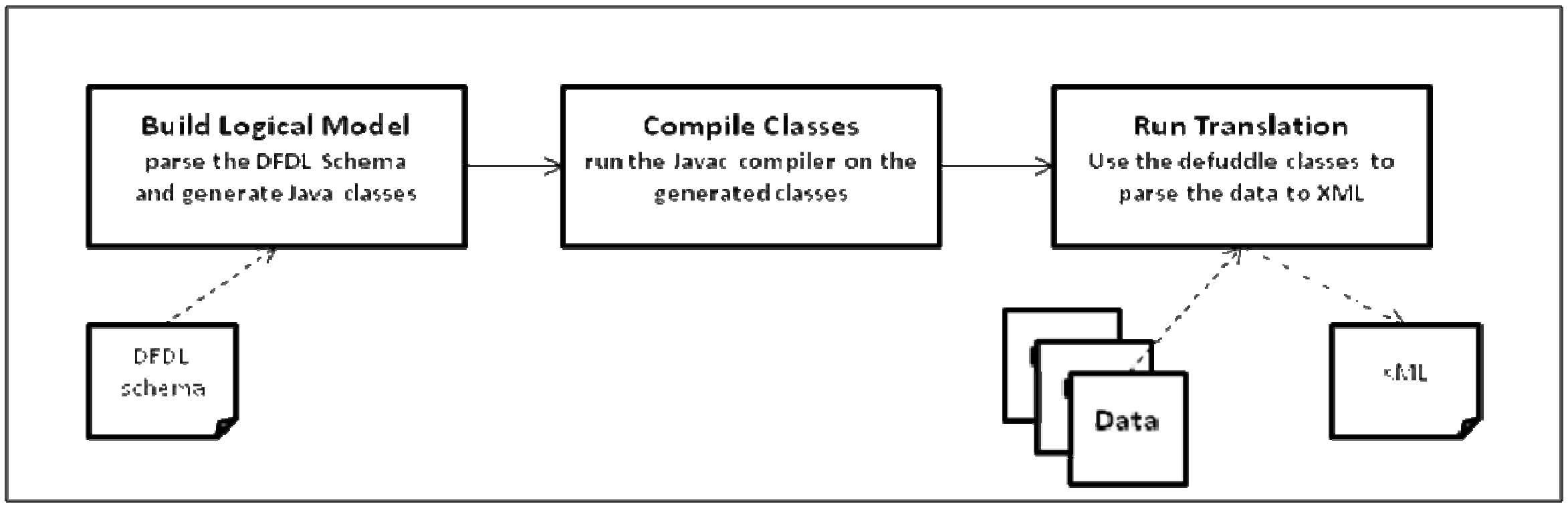

**Figure 1. Basic operation of the Defuddle parser.**

it is physically represented.

Independently of the DFDL/Defuddle efforts, the emphasis in the broad community in defining standard logical models for data is evolving from thinking in terms of local, hierarchical schema to more global, network-oriented models, i.e. through the use of semantic web representations such as Resource Description Framework (RDF) and Web Ontology Language (OWL). As noted in the Introduction, there are significant advantages to be had in associating data within a file with standardized community vocabularies, reference knowledge about the world, and additional descriptive and data processing (provenance) information. This additional context can provide 'missing' detail that is not directly available from the format (e.g. unit information may be associated with software producing the file rather than the format itself) and provides an enhanced basis for preserving data semantics.

In theory, it might be possible to construct a direct analog of DFDL that would be associated with RDF Schema or OWL ontologies and develop a parser analogous to Defuddle to generate classes capable of reading ASCII/binary and producing RDF output. However, RDF and OWL are not structural descriptions of data, so such a markup language would require development of a mapping language between data and relations, which is substantially different and more difficult than the problem solved by DFDL (mapping data structures to semantic relations vs. mapping sequential data structures into XML data structures). Given this difficulty, it is not surprising that the equivalent of the JAXB binding and robust tooling for such a standard do not yet exist. Hence such a route would involve a prohibitive level of effort.

An alternate route extends the mapping from ASCII/binary to XML to a second mapping from XML to RDF. RDF can be represented as an XML dialect and standard XSLT mapping mechanisms can convert between arbitrary XML logical models and XML-encoded RDF. Our approach is such a two-step process: the standard DFDL to generate XML, and a second phase to extract semantic descriptions from the XML. The second phase uses the GRDDL standard to generate RDF from the XML.

The Gleaning Resource Descriptions from Dialects of Languages (GRDDL) standard (http://www.w3.org/TR/grddl/) is a convenient mechanism for associating such XSLT transforms with XML schema and XML instance documents. GRDDL is a standard markup that specifies transformations (usually XSLT) that extract RDF from the XML. While the XSLT or other extraction can be applied in many ways, GRDDL provides a standard mechanism for *declaring these transformations within the XML schema and/or XML instance documents*. Thus, it is quite consistent with the declarative approach of DFDL. The GRDDL Specification [3, 16] and related documents are "W3C Recommendations." A formal test suite is defined (see [12]) and at least three implementations exist (see [4]):

1. a web service (http://www.w3.org/2007/08/grddl/)
2. a python GRDDL.py (http://www.w3.org/2001/sw/grddl-wg/td/GRDDL.py),
3. a Jena plug-in (in Jena 2.5.1+) (http://jena.sourceforge.net/grddl/).

This approach is guided by practical considerations of the available code base as well as the desire to keep both XML and RDF output capabilities. Furthermore, we build on the strength of each language—XML for structural description of data and RDF for description of relations. This approach is explained in detail below.

# 3. Current Work

Our current work builds on the Defuddle parser originally developed at PNNL [15]. The initial implementation by PNNL was released as open source software but was developed concurrently with efforts to create a DFDL test suite and did not receive regular maintenance. Hence, the effort at NCSA has begun with an effort to update the core Defuddle parser, assess performance and correctness, and develop a robust software engineering environment and build and test environment. NCSA is also extending Defuddle to support semantic data models, e,g, extending it to extract RDF descriptions of structures and relations in the data. The extended Defuddle parser is then being integrated into iRODS as a microservice and will be linked with SHAMAN's Multivalent Browser (MVB [18]) as part of the overall SHAMAN effort. These development areas are discussed in more detail in the following subsections.

## 3.1. DFDL and the Defuddle Parser

Defuddle has been migrated to the latest version of JaxMe[2], the first significant upgrade to Defuddle since 2006. The port was substantially more complex than originally anticipated, requiring merging changes in approximately fifty files, to adapt to extensive, fundamental changes to the JaxMe internals. The new version has been tested against the same example set as the earlier release and found to reproduce the results and will thus become the next official version of Defuddle. In fact, this regression testing has identified several issues that appear to be long standing (i.e., they were present in the previously released code). These issues are being investigated and will be fixed in a future release. The new implementation of Defuddle has also been updated to use Maven for software build management, increasing automation in managing future software updates.

Based on this version, NCSA has begun exploring uses of Defuddle to interpret files as part of other projects. These efforts include initial work to extract metadata from 3-D file formats in concert with Peter Bajcsy's effort within the same "Innovative Systems and Software: Applications to NARA Research Problems" project and work to extract common information from multiple related ASCII formats in the area of infectious disease monitoring (in collaboration with Ian Brooks at NCSA). While these activities are preliminary, they represent a move from proof of concept to a phase of assessing Defuddle in real-world use cases.

**Technical Detail**

As noted above, Defuddle is implemented as an extension of a Java/XML binding compiler (JaxMe) based on the Java Architecture for XML Binding (JAXB) specification. JAXB provides a convenient way to use XML Schema to automatically parse XML instance documents into Java classes corresponding to that schema From a design standpoint, this solution provides an off-the-shelf ingestion engine for the logical model (XML Schema), a dynamic logical model compiler (Java classes), and an XML document generator for streaming data from the classes to XML.

Figure 1 illustrates the conceptual design of the Defuddle parser. At run-time, the schema is ingested and processed to generate Java classes representing components of the logical model. These classes are then compiled using the standard Java compiler. The translation of the input data source(s) is then initiated using the JAXB XML marshaller. As the java objects are streamed by the parser, the logical model is formed by loading the required values from the data file into

---

[2] JaxMe is the Apache open source software implementation of the Java for XML Binding (JAXB) specification (http://ws.apache.org/jaxme/). JaxMe and JAXB are used interchangeably throughout this document.

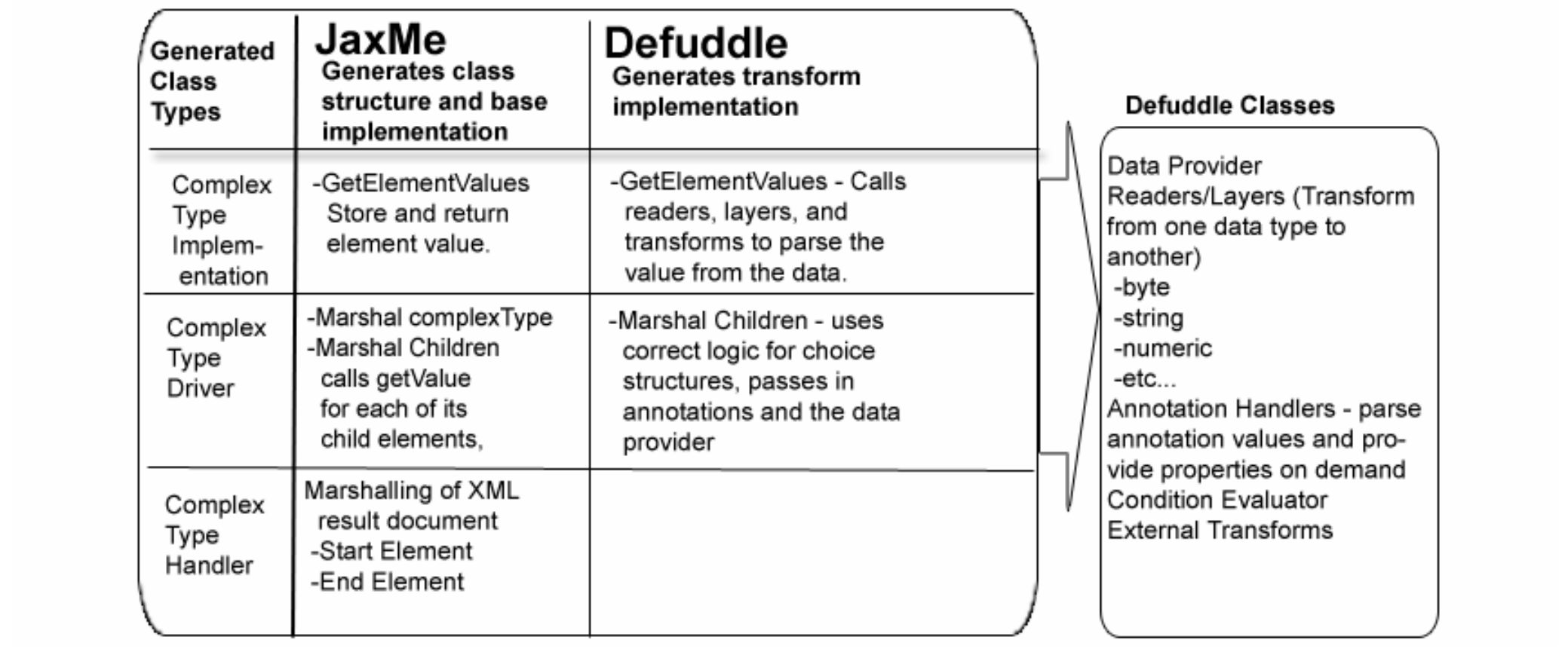

**Figure 2 Defuddle Extensions to JaxMe (adapted from Talbott [15].**

Java objects corresponding to the logical model. This set of Java objects represents an XML Infoset model that can then be streamed to an output that can then be processed by standard XML tools. By default, Defuddle is configured as a command line tool that accepts source files and format description files, performs the compilation and execution steps, and outputs XML documents containing data formatted in the defined logical model. The class generation process is performed automatically before translations are performed, but the compiled classes can be cached to improve performance on subsequent runs. Further, it would also be possible to expose Defuddle as a library that could be used within a Java program. Such a library could accept data streams directly (rather than files) and could output the Java object model rather than an XML document.

Figure 2 illustrates the types of classes generated by JaxMe and the Defuddle extensions that implement various features. Each complex type is represented by three classes: the type implementation, type driver, and type handler. While the `complexType` implementation classes provide information to parse individual values, the parser needs additional information to understand the structure of the data. This includes the order of the elements, the location of sequences, and the type of data to be marshaled. This functionality is found in the `complexType` drivers. Within the type implementation, values such as elements and attributes are accessed from the data stream using `get<Name>` methods. Vanilla JaxMe-generated type implementations store and return the values, and Defuddle adds content to these ‘get’ methods, parses the required data using the annotation handlers and data provider along with other built-in Defuddle classes, such as the type readers, condition evaluators, and external transforms.

The basic ordering and ‘get’ calls are generated by JaxMe. Defuddle extensions check for layers, hidden elements, and control sequences of unknown lengths. They also choose the correct element in conditional statements and pass in the data provider and annotation values. The third kind of generated class is the `complexTypeHandler`; these classes are generated almost entirely by the JaxMe generator and chiefly control the marshalling of the classes to XML, ensuring the correct state when starting/closing elements and writing data.

Along with the JaxMe extensions, Defuddle contains additional classes to aid in parsing as shown in the right side of Figure 2. Various readers are used for converting values from one type to another, for example from byte to string, from string to multiple strings, and from string to integer. Each of these readers use the annotation properties specified in the schema. Defuddle also uses a data provider to retrieve from the data stream, referencing other values in the schema, or for handling multiple input sources. When evaluating conditions, the annotations are passed on to a condition evaluator to determine the correct element to read. Defuddle also supports the calling of external transformations and integrating the results into the Defuddle transformation via `ExternalTransform` classes.

## 3.2. Semantic Extensions to Defuddle

As discussed above, while the XML Schema language is well suited for describing the layout of data (the "syntax"), interoperability and robust archiving require semantic markup as well. The DFDL model converts data into XML; or, alternatively, extracts XML from the data; in order to gain the advantages of the XML standard for data representation. Our goal in this task has been to extend this model to extract descriptions of the structure and relations in the data into standard semantic web languages (the Resource Description Framework (RDF) (http://www.w3.org/TR/rdf-concepts/) and the Web Ontology Language (OWL) (http://www.w3.org/TR/owl-features/).

Our approach is a two-step process, the standard DFDL processing within Defuddle to generate XML, and a second phase to extract semantic descriptions from the XML (see Figure 3). The second phase uses GRDDL as a standard mechanism for *declaring these transformations*.

After an initial evaluation of the existing GRRDL implementations (listed in the Background section), we chose to work with the GRRDL.py implementation given its reasonable level of maturity and command-line orientation. It will be possible to create alternative modules that use the Jena plug-in or other implementation in the future.

Conceptually then, our semantically extended Defuddle focuses on a standard XML schema, decorated by two types of annotations:

1. DFDL (with Defuddle extensions) to describe the original data
2. GRDDL plus XSLT describe structures and relations in the transformed data

The Defuddle parser simply applies another step, analogous to executing the DFDL-based

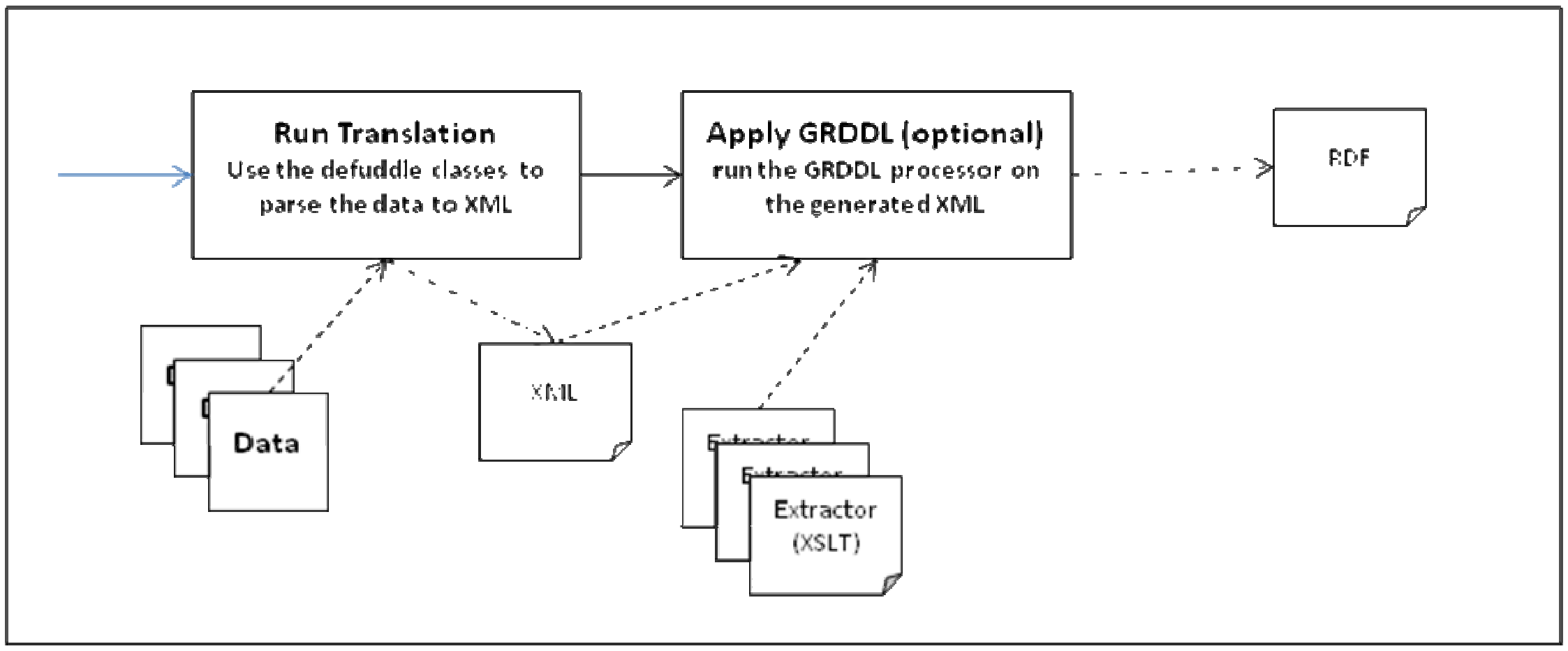

**Figure 3. Basic operation of the GRDDL extension to Defuddle.**

translation code, to invoke a command–line GRDDL.py step. The Defuddle parser generates two outputs:

1. XML which conforms to the schema
2. RDF which describes the data

In practice, we have added a third step in which GRDDL statements to the XML output from Defuddle which then define the processing that will occur in the GRDDL.py stage. This enables the user to specify a set of GRDDL transformations to execute for a given file. A second use of GRDDL is for the XML schema to specify transforms that will be applied to all files. Both schemes can be used in the same transformation: the latter defines processing for any file that uses the schema; the former can define specific transforms for individual files. Ultimately there may be use cases requiring both models to be supported, in which conversion to RDF is seen as an integral part of DFDL transformation and/or a separate step that may be applicable to the logical model (regardless of which source format the data come from) or only within certain communities (e.g., reasoning that while the logical model is a property of the format, the appropriate global vocabulary is an independent choice made by communities and may be more closely associated with common logical substructures across models).

RDF can also be used to retain the provenance of this pipeline [6, 7]. The result would be data in standard formats (RDF and XML) with additional references to standard schemas and ontologies. The RDF and XML can be further processed to transform the data structures and relations into whatever abstract forms are needed. In particular, the data can be transformed into data structures needed for analysis and visualization, (e.g., by the MVB [18]). This work is based on the intuition that, for any given case, it is possible create an "ontology" of the operations for structures and relations. Such an ontology would build on standard concepts of data structures and relations, to implement descriptions of domain specific structures and relations. For example, a hydrology dataset could be described in terms of domain-specific objects that are composed of a specific collection and arrangement of generic arrays, strings, tuples, and so on.

DFDL augmented with RDF provide general mechanisms for processing these descriptions. RDF can store representations of operations in portable and machine readable forms, and RDF would be amenable to automated reasoning (e.g., to guess appropriate default actions). However, such automated reasoning techniques cannot be created without a working conceptual model. These technologies do not define what the abstract operations are, nor is there any general model for them. Our work will show how to implement examples of these mappings derived from our collaborators.

### Technical Detail

The GRDDL standard defines assertions to declare the GRDDL name space and then list one or more "transformations" to be applied ([2], http://www.w3.org/2001/sw/grddl-wg/). The transformations are URIs that point to a description of a transformation. For our purposes, the transformations are XSL stylesheets that process the XML output of Defuddle and generate RDF triples. For example, in an XML file with the top level element "workflow", to apply a GRDDL transformation called "vistrails2rdf.xsl" (which we use to convert provenance created by Utah's Vistrails visualization software to an RDF binding of the Open Provenance Model), the XML would be annotated in the following manner:

```
<workflow id="120" name="part1"
xmlns:grddl="http://www.w3.org/2003/g/data-view#"
grddl:transformation="http://vesta.ncsa.uiuc.edu/GRDDL/xsl/vistrails2rd
f.xsl">
```

In our initial approach, the Defuddle software controls insertion of the required text in the XML it generates. This approach is simple and is known to work with current GRDDL engines. Since Defuddle controls what is written in the XML, the GRDDL can always be inserted. The initial implementation inserts the GRDDL assertions into the output model–a step managed as a command-line option-rather than via programmatic insertion which would require a registry of logical model to desired RDF/OWL vocabularies.

There are several implementations of the GRDDL standard available as noted above. The two most complete are GRDDL.py (http://www.w3.org/2001/sw/grddl-wg/td/GRDDL.py) and the Jena Plugin (http://jena.sourceforge.net/grddl/). The former requires Python but is self-contained. The latter is part of Jena and thus is linked directly from Java. Our initial implementation uses GRDDL.py because it is more convenient than the Jena plug-in. Future implementations could make the choice of GRDDL engine a configuration option. In any case, the input files, DFDL and GRDDL markup, and GRDDL transforms are the same no matter which engine is used. The engines differ only in how they are invoked, how the output is provided, and potentially on robustness and performance.

GRDDL transforms are XSLT that reads XML and generates RDF. GRDDL does not specify the content of transforms except that they generate RDF. Note that any number of transforms may be applied to a given XML document, thus making it possible to consider each logical type/substructure in a model independently and to reuse XSLT files generated in the broad community. For example, transforms are available which process Friend-of-a-friend (FOAF) records [1] which can be used to extract relations from any XML that contains FOAF records. Such a transform could be used in combination with those that extract bibliographic information from Dublin Core, transforms to scientific unit ontologies, and so on.

In order to extract novel structures and relations of interest with Defuddle, it will be necessary to create one or more XSL transforms. Once transforms are created, they can be published at well-known URLs and used by any GRDDL engine, and hence may be used as community resources independent of DFDL/Defuddle. One of the central purposes of the GRDDL standard is to provide a declarative mechanism to document which transforms can be used for a given XML schema or instance and create a framework for community sharing.

In the current implementation, the selection of which transforms to apply is a decision made at runtime by the person executing the Defuddle command-line tool (or, similarly, an iRODS microservice as discussed below). As discussed above, alternate models for managing this selection include making the GRDDL annotations part of the DFDL file (given that they are allowed by XML Schema, there is no change to the DFDL language needed to support them) and/or developing a registry/recognition mechanism to dynamically choose relevant transforms given the XML output being generated. To move beyond a proof-of-principle implementation of a semantic Defuddle will require consideration of these issues, development of an online respository/registry of DFDL and GRDDL-style XSLT translators, and creation of appropriate test examples. Additional enhancements to the Defuddle command-line tool to support additional properties controlling input and output locations and potentially translator options may also be needed.

The RDF output from the semantically enhanced Defuddle can be considered both as a file and as a collection of triples. When more than one transformation is applied, all the triples are aggregated into a single collection. GRDDL engines have different options for how the RDF is produced, i.e., how the triples are serialized. The initial semantic Defuddle implementation generates XML-encoded RDF and simply exports that as a file. However, use cases such as those being considered in SHAMAN will derive value from managing Defuddle-derived triples as part of a larger graph of information including provenance, domain ontologies, and social network information. To support this, the RDF triples would need to be ingested into some form of triple store where they can be queried and used in subsequent logical inference, rule-based, and other processes. Given NCSA's extensive work in developing semantic content middleware (Tupelo [5], http://www.tupeloproject.org), it would be natural to store Defuddle's output via this middleware. Preliminary work in this direction has begun as part of a Google Summer Of Code Student Fellowship in which Yigang Zhou will be implementing a mechanism (as a Tupelo Context) to invoke Defuddle as part of a file upload operation in Tupelo and to store the output as metadata or new content linked to the original data via provenance information (see http://socghop.appspot.com/student_project/show/google/gsoc2009/ncsa/t124022775909).

## 3.3. Archiving Experiment Test bed

Part of the EU SHAMAN project ([14]) is an end-to-end demonstration is being developed by U. Liverpool and DICE, which will include a complete archiving environment built using iRODS data grid technology [10], including discovery and access (i.e., use after storage). The overall conceptual scheme is suggested in [9, 17], which describes "characterizing preservation processes", including not only the bits, but also logical structures and relationships. While the MVB is designed for documents and images, it needs to be extended to deal with scientific or structured data [18]. The SHAMAN prototype plans to create the preservation metadata in the Open Archives Initiative Object Reuse and Exchange standard (OAI-ORE), specifically, using the RDF binding [8].

A prototype Defuddle microservice is being developed as a method to integrate Defuddle into this prototype. Defuddle provides a technology by which other forms of data can be brought into this system by extracting XML and RDF representations of the non-text objects. There are a great variety of such data; with semantics at various levels of detail and specificity, Defuddle can apply alternative transformations from different logical perspectives, generating RDF that can be used with other RDF metadata. Thus, additional information generated through these uses of Defuddle can be captured for long-term use. However, there could be additional value in coupling Defuddle+GRRDL with a database-style semantic store, i.e. one that supports querying (i.e. using SPARQL [13]), inference, rule-based processing, etc. This would allow the direct outputs from Defuddle to be further processed in the overall context of collections, projects, and community knowledge bases to generate additional metadata.

To support this effort, Defuddle has been wrapped as an iRODS microservice. Such a microservice should enable Defuddle to be executed on a variety of input data, with the results passed to other microservices and/or stored to one or more repositories. In some cases, the XML will be used by other services, so it will be a new derived data product in its own right. Similarly, the RDF metadata may be used by other services as content (as a file) or as metadata to be combined with other RDF.

A basic use case is:

1. A file is uploaded to iRODS.
2. iRODS rules determine that Defuddle should be executed.
3. Defuddle microservice executes using appropriate schema and GRDDL transforms.
4. XML output is passed to other microservice or stored to repository.
5. RDF output is passed to other microservice and/or stored in triple store.
6. Return status and errors.

NCSA has implemented an (alpha-level) microservice that can ingest data in various formats and use the Defuddle parser to convert it into standard XML, with the resulting XML made available for storage or for passing to other microservices for further processing. The microservice supports both standard DFDL processing as well as the GRDDL phase to extract RDF from the XML. It is anticipated that this proof-of-concept will help spur dialog within SHAMAN to identify specific use cases and associated logical models that will provide the basis for an integrated demonstration with SHAMAN and provide additional information about the level of status and error reporting that will ultimately be required. To date, it appears that the effort within SHAMAN is primarily focused on core ingest and storage with limited emphasis on defining the integration of MVB and iRODS. As this connection is solidified it will help clarify the logical models that will be required from Defuddle.

### Technical Detail

The Defuddle microservice is implemented as a C plug-in to the iRods framework. The C module calls Java to execute the Defuddle command line client. The microservice is invoked with arguments that parallel the Defuddle command line program (Figure 4):

- data file
- schema
- optional XSL
- GRDDL XSL
- output file

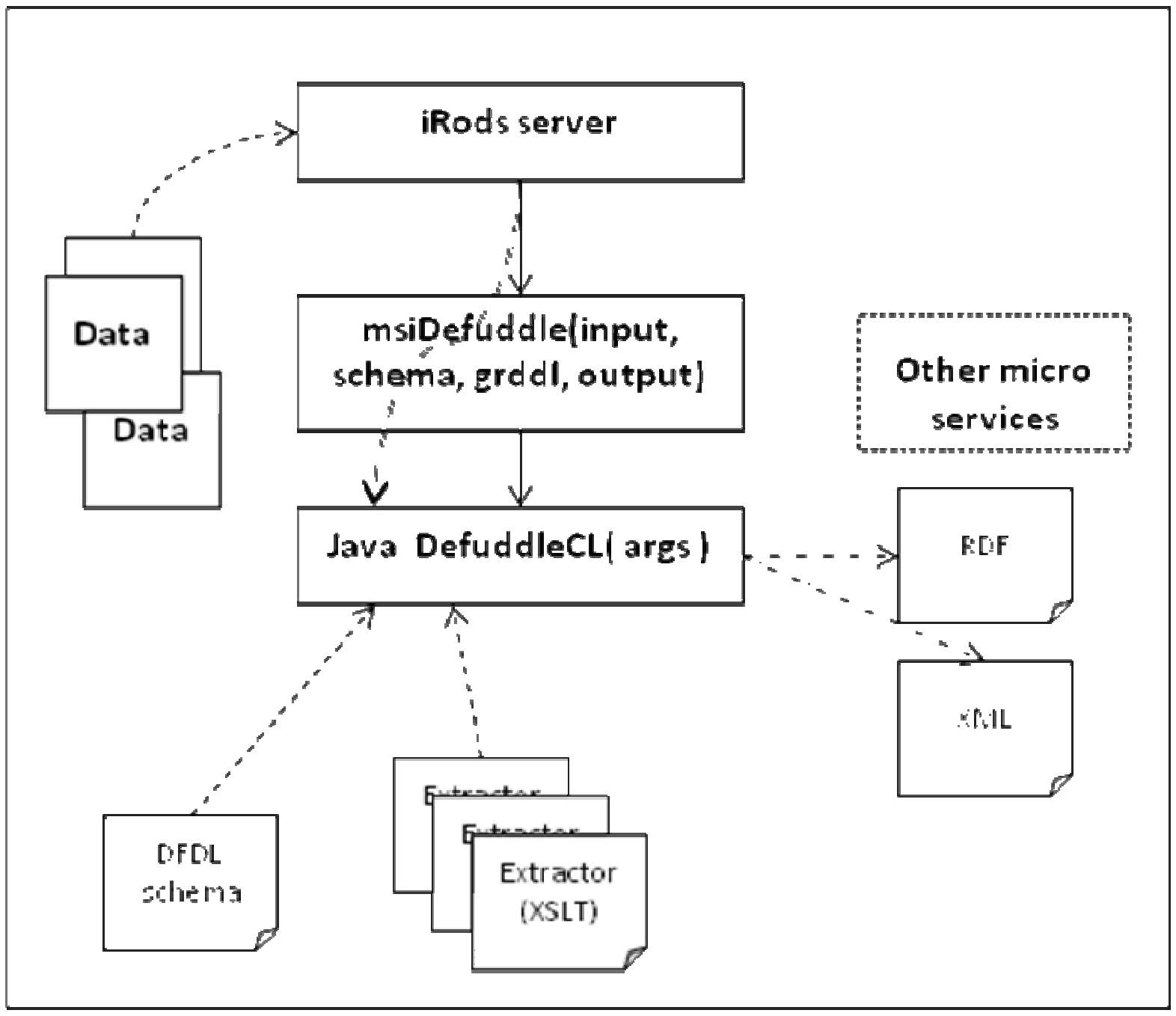

**Figure 4. Basic operation of the Defuddle microservice.**

The microservice collects the data, schema, and XSL files and then invokes the standard Defuddle program. The output is pushed into the iRods repository. The microservice is invoked by iRODS rules. Figure 5 shows an example rule.

```
myDefuddleRule||msiCallDefuddle(*schema,*data,*data.xml)|nop
*schema=/tempZone/home/tester/A2.xsd%*data=
/tempZone/home/tester/asciint.txt
*null
```

**Figure 5. Example iRODS rule to invoke the Defuddle microservice.**

The initial microservice has served to identify a number of basic issues which will be investigated, including:

- How to match input file with appropriate schema
- How to manage collections of schemas and transforms
- How to buffer data, minimizing copies
- How to manage output, XML and RDF

***Matching inputs to schema.*** The basic trigger for the operation of the microservice occurs when a file is 'put' into a repository. The idea is for iRODS to invoke Defuddle to convert the data to XML and extract metadata into RDF, which would be deposited into an appropriate service. However, the Defuddle parser itself must be told what schema to use. Nominally, this could be done through an iRODS rule (e.g., to detect a MIME type or other identifying information) and determine an appropriate DFDL schema to apply, but it also may be useful to have the microservice manage its own registry of MIME type to DFDL file mappings. Additional work and discussion with the iRODS team and SHAMAN will be needed to assess the pros and cons of these approaches and to development example rules and/or supporting microservices required to implement the most promising solution.

***Managing collections of schemas and transforms.*** The iRODS system is designed to manage collections which may be owned by different communities and, as a result, subject to different rule sets. The set of transforms used by Defuddle and their mapping to MIME types will potentially be scoped by collection as well. Further, since the transforms and mapping information could themselves be considered a collection, the potential exists for iRODS to manage them as a collection via existing mechanisms. For this to be effective, the association between different Defuddle transform collections/subcollections and the domain data collections they are used with must be maintained. At this stage, NCSA does not have a deep enough understanding of iRODS to access the practicality of this option or to assess the potential impact on community use of the transforms outside iRODS (communities may wish to invoke DFDL-based transformations as part of scientific workflows or other processes and thus may want the transforms and MIME type mapping information available externally). Steps such as the GRDDL transformation to RDF could potentially require valid URLs for DFDL schema as well. It should be possible for the microservice to use external schemas and transforms (e.g., from a community portal); however, because of limitations of implementations or networks, or simply in order to meet preservation requirements, it may be necessary to copy these files to working storage on the iRODS server. When this is done, some mechanism will be needed to maintain consistency with the original sources. Thus, further discussion and potentially prototyping efforts will be required to assess the best mechanism for supporting collection-scoped use of DFDL/Defuddle within iRODS and how to make that most compatible with broader use of DFDL/Defuddle in other pieces of infrastructure.

***Minimizing copies.*** The current implementation of the Defuddle parser copies the input file to a working area, generates output to a working area, and then pushes the result out. This simple design copies data to and from disk several times, which is a potential performance issue. (This is also an issue in memory as will be discussed further in the Performance section). In any case, these operations need to be managed: space must be allocated and cleaned up, users must be isolated from each other, etc.

When embedded in a microservice, these operations must happen within the context of an iRODS server. In some cases, the data may be coming in as a stream from the network; in other cases, Defuddle may be applied to an object stored in the repository. The output will presumably be stored into the repository. It will be important in moving from demonstration to operations in order to understand the best ways to manage scratch space, how to retrieve and store to permanent storage, and how to interact with other services within the iRODS server.

***Managing output, XML and RDF.*** A Defuddle microservice creates XML and/or RDF which may be stored in the repository and/or sent to other repositories and services. The default output of the current Defuddle microservice is a set of working files, but the service could potentially return streams or sets of XML key/value metadata or RDF triple collections directly given appropriate interfaces to iRODS and with other microservices. Regardless of format, the output objects need to be assigned identifiers, and their association with the original dataset and transform should be maintained. iRODS manages such information, but it may be useful for the microservice to record such provenance information, in more detail or in standard community formats such as the proposed Open Provenance Model [11], as well. Requirements in this area, and the extent of capabilities for iRODS to manage detailed provenance from microservices and to export that information to other microservices and applications, are not yet clear, and hence additional discussion and development effort are likely to be required in this area.

# 4. A Defuddle Example

This section presents a simple example of how Defuddle is used, including a simple example of how semantic relations are extracted.

Figure 6 shows a simple data file, which has two lines of metadata followed by rows of data. Each row has 10 numbers separated by commas and ending with an end of line. The data is stored in ASCII.

```
Creator: NCSA
Date: Mon Feb 23 15:20:47 CST 2009
0,1,2,3,4,5,6,7,8,9
1,2,3,4,5,6,7,8,9,10
2,3,4,5,6,7,8,9,10,11
3,4,5,6,7,8,9,10,11,12
4,5,6,7,8,9,10,11,12,13
5,6,7,8,9,10,11,12,13,14
6,7,8,9,10,11,12,13,14,15
7,8,9,10,11,12,13,14,15,16
8,9,10,11,12,13,14,15,16,17
9,10,11,12,13,14,15,16,17,18
```

**Figure 6. Simple example of input data**

DFDL/Defuddle will be used to convert this data into a logical model defined in terms of structured XML specified in a schema document. Figure 7 shows the top level structure of the desired output for this example: a block of metadata followed by a block of data.

```
<xs:element name="table" type="SimpleTable">
</xs:element>

  <xs:complexType name ="SimpleTable">
      <xs:sequence>
            <xs:element name="hdrblock" type="header" />
            <xs:element name="datablock" type="Row"
maxOccurs="unbounded"/>
      </xs:sequence>
  </xs:complexType>
```

**Figure 7. The top level structure described in XML Schema.**

Figure 8 defines the layout of the metadata: an 'Author' field and a 'CreationDate' field. Figure 9 shows the layout of the data rows, each number is in its own XML element. The complete schema is given in Appendix A. Note that through the use of XML `ComplexTypes`, the logical header block and row-oriented data structure could potentially be reused in multiple DFDL descriptions (e.g., for files which may contain additional data structures but share logical components with this example).

```
<xs:complexType name="header">
      <xs:sequence>
            <xs:annotation>
                  <xs:appinfo>
                        <dfdl:dataFormat>
                              <dfdl:repType>text</dfdl:repType>
                        </dfdl:dataFormat>
                  </xs:appinfo>
            </xs:annotation>
            <xs:element name="Author" type="xs:string" >
                  <xs:annotation>
                        <xs:appinfo>
                           <dfdl:dataFormat>
                              <dfdl:ignore>Creator:\\s</dfdl:ignore>
                              <dfdl:terminator
              kind="regexp or
string">\\r\\n|[\\r\\n]</dfdl:terminator>
                              <dfdl:charset>US-ASCII</dfdl:charset>
                           </dfdl:dataFormat>
                        </xs:appinfo>
                  </xs:annotation>
            </xs:element>
            <xs:element name="CreationDate" type="xs:string" >
                  <xs:annotation>
                        <xs:appinfo>
                           <dfdl:dataFormat >
                                 <dfdl:ignore>Date:\\s</dfdl:ignore>
                                 <dfdl:terminator
           kind="regexp or string">\\r\\n|[\\r\\n]</dfdl:terminator>
                                 <dfdl:charset>US-ASCII</dfdl:charset>
                           </dfdl:dataFormat>
                        </xs:appinfo>
                  </xs:annotation>
            </xs:element>
      </xs:sequence>
</xs:complexType>
```

**Figure 8. XML Schema describing the logical format of the header. DFDL annotations (Bold) describe the input file, i.e., how to read the data into the XML structures.**

The logical structure of the schema is mapped to the input file (Figure 6) with DFDL annotations in the schema (shown in bold in Figure 7-9). The 'repType' (representation type) indicates that the data is text. The 'separator' and 'terminator' define patterns that separate the elements and blocks respectively, the 'ignore' tag defines data to be skipped. As shown in Figure 8 and Figure 9, separators can be specified via regular expressions which can include 'or' clauses to handle format variations (e.g. files that may differ in the character sequence used for separation but are otherwise identical in structure). Other aspects of DFDL including 'if' and 'loop' logic (not included in this example) broaden the types of format variation that can be managed by a single DFDL description file. The annotations in the example also specify the base (10) and encoding (US-ASCII).

```
<xs:complexType name="Row">
      <xs:sequence>
            <xs:annotation>
                  <xs:appinfo>
                        <dfdl:dataFormat>
                              <dfdl:repType>text</dfdl:repType>
                        </dfdl:dataFormat>
                  </xs:appinfo>
            </xs:annotation>
            <xs:element name="item" type="xs:int"
                    minOccurs="10" maxOccurs="10">
                  <xs:annotation>
                        <xs:appinfo>
                              <dfdl:dataFormat xmlns:dfdl="DFDL">
                                    <dfdl:separator kind=
        "regexp or string">,\\r\\n|,[\\r\\n]|,|\\r\\n|[\\r\\n]
                                    </dfdl:separator>
                                    <dfdl:base>10</dfdl:base>
                                    <dfdl:charset>US-
ASCII</dfdl:charset>
                              </dfdl:dataFormat>
                        </xs:appinfo>
                  </xs:annotation>
            </xs:element>
      </xs:sequence>
</xs:complexType>
```

**Figure 9 XML Schema describing logical type of the data block. DFDL annotations (Bold) describe how to read the data.**

```
<?xml version='1.0' encoding='UTF-8'?>
<dataset:table xmlns:dataset="/URI/of/schema">
  <dataset:hdrblock>
    <dfdl:Author>NCSA</dataset:Author>
    <dataset:CreationDate>Mon Feb 23 15:20:47 CST
2009</dataset:CreationDate>
  </dataset:hdrblock>
  <dataset:datablock>
    <dataset:item>0</dataset:item>
    <dataset:item>1</dataset:item>
    <dataset:item>2</dataset:item>
    <dataset:item>3</dataset:item>
    <dataset:item>4</dataset:item>
    <dataset:item>5</dataset:item>
    <dataset:item>6</dataset:item>
    <dataset:item>7</dataset:item>
    <dataset:item>8</dataset:item>
    <dataset:item>9</dataset:item>
  </dataset:datablock>
  <dataset:datablock>
   […]
  <dataset:datablock/>
</dataset:table>
```

**Figure 10. XML output (partial). This follows the logical layout of the schema, populated by the contents of the original data.**

The Defuddle parser reads this annotated schema to create a parser. The parser includes code to read the text and convert it to the required numeric value, which is included in a DOM tree.

The compiled parser may be executed for any data file that matches the specification. The parser reads the data and creates an XML data model that conforms to the schema. Figure 10 shows partial output of this phase for the example data in Figure 6. The data elements match the input file, and the structure matches the logical schema.

The second phase in our semantically enhanced Defuddle parser applies GRDDL to extract RDF from the XML. For example, the metadata in the header (the author and date) could be extracted to generate RDF records conforming to a standard such as Dublin Core. This is done by creating an XSL style sheet that can be applied to the XML. Figure 11 shows a simple XSL that pulls out the 'Author' and 'CreationDate' from the XML and generates Dublin Core-conformant RDF. In this case, the XML fields are directly related to the RDF ("Author" => dc:creator, "CreationDate" => dc:date), but this need not be so. The GRDDL can change the names of items or perform any other transformation allowed by XSLT to create RDF.

When GRDDL is selected, the Defuddle parser inserts GRDDL markup into the XML output. Figure 12 shows the necessary markup added to the example from Figure 10. The GRDDL.py program is called on the generated XML. The result is a set of RDF triples encoded in XML. Figure 13 shows an example of the output produced by GRDDL using the XSL in Figure 11 on the XML in Figure 13.

```
<?xml version="1.0" encoding="utf-8"?>
<!DOCTYPE xsl:stylesheet [
         <!ENTITY rdf 'http://www.w3.org/1999/02/22-rdf-syntax-ns#'>
       <!ENTITY rdfs 'http://www.w3.org/2000/01/rdf-schema#'>
         <!ENTITY dataset 'Dataset'>
         <!ENTITY xbsdfdl 'http://ncsa.uiuc.edu/dataset#'>
         <!ENTITY dfdl 'DFDL'>
         <!ENTITY absdfdl 'http://ncsa.uiuc.edu/DFDL#'>
]>
<xsl:stylesheet xmlns:xsl="http://www.w3.org/1999/XSL/Transform"
                xmlns:rdf="&rdf;" xmlns:rdfs="&rdfs;"
                xmlns:dfdl="&dfdl;" xmlns:absdfdl="&absdfdl;"
                xmlns:dc="http://purl.org/dc/elements/1.1/"
                version="1.0">
  <xsl:output method="xml" version="1.0" encoding="utf-8"
indent="yes"/>
  <xsl:template match="/dfdl:table">
    <rdf:RDF>
      <rdf:Description>
        <rdf:type rdf:resource="&absdfdl;table"/>
        <dc:creator><xsl:value-of
                  select="dfdl:hdrblock/dfdl:Author"/></dc:creator>
        <dc:date><xsl:value-of
                 select="dfdl:hdrblock/dfdl:CreationDate"/></dc:date>
      </rdf:Description>
    </rdf:RDF>
  </xsl:template>
</xsl:stylesheet>
```

**Figure 11. XSL to create RDF triples expressing the descriptive metadata.**

```
<?xml version="1.0" encoding="UTF-8"?>
<dataset:table xmlns:dataset="Dataset"
xmlns:ns0="http://www.w3.org/2003/g/data-view#"
ns0:transformation="examples/schemas/SimpleCSV.xsl" >
<dataset:hdrblock>
    <dataset:Author>NCSA</dataset:Author>
    <dataset:CreationDate>Mon Feb 23 15:20:47 CST 2009
</dataset:CreationDate>
  </dataset:hdrblock>
<dataset:datablock>
    <dataset:item>0</dataset:item>
    <dataset:item>1</dataset:item>
    <dataset:item>2</dataset:item>
    <dataset:item>3</dataset:item>
    <dataset:item>4</dataset:item>
    <dataset:item>5</dataset:item>
    <dataset:item>6</dataset:item>
    <dataset:item>7</dataset:item>
    <dataset:item>8</dataset:item>
    <dataset:item>9</dataset:item>
  </dataset:datablock>
<dataset:datablock>
    <dataset:item>1</dataset:item>
    …
  </dataset:datablock>
<dataset:datablock/>
</dataset:table>
```

**Figure 12. Partial XML output with GRDDL annotation inserted (Bold). Compare to Figure 8.**

```
<?xml version="1.0" encoding="UTF-8"?>
<rdf:RDF
   xmlns:dc="http://purl.org/dc/elements/1.1/"
   xmlns:rdf="http://www.w3.org/1999/02/22-rdf-syntax-ns#" >
  <rdf:Description rdf:nodeID="OaJonZub4">
    <dc:creator>NCSA</dc:creator>
    <dc:date>Mon Feb 23 15:20:47 CST 2009</dc:date>
    <rdf:type rdf:resource="http://ncsa.uiuc.edu/dataset#table"/>
  </rdf:Description>
</rdf:RDF>
```

**Figure 13. Example RDF output by GRDDL.**

This example demonstrates the basic function of the GRDDL extraction. The value of the RDF generated depends on the transformations that are defined. As noted in the previous section, Defuddle may eventually need to generate RDF that identifies the original and transformed data objects (i.e., the files in Figure 6 and Figure 10) with valid URIs, as well as the provenance of these files, including the transforms specified in Figure 7-9 and Figure 11. As discussed above, this step requires cooperation with wider infrastructure, such as Tupelo (http://tupeloproject.ncsa.uiuc.edu/) or, when running as a microservice, other iRODS services, and such provenance outputs could potentially be managed differently than RDF generated directly by the GRDDL transforms.

# 5. Performance Investigations

The performance of the Defuddle parser was examined using the ‘SimpleCSV’ dataset illustrated above, expanded by adding more data in the same pattern (i.e., more rows of numbers). As might be expected, the most computationally intensive aspect of Defuddle processing is the parsing of the data file to produce XML.[3] The analysis of this example provides some general conclusions about the performance of Defuddle as discussed below. However, in general, the performance of Defuddle (and any DFDL parser) will be highly dependent upon the complexity of the data, and of the logical model, and of the mapping between them. Thus, the quantitative results presented here should not be interpreted as being applicable across file formats.

Overall, the XML output for this example was about eight times the size (bytes) of the input data. This is about what would be expected just from XML encoding. Given the close match between the ASCII format and logical model, no other sources of file size increase are expected. (However, consider an example where data has been run length-encoded – the XML output could be significantly larger if the logical model eliminates the compression).

The principal finding from the performance analysis is that the performance depends fairly linearly on the size of the input file and does have a dependence on the logical structure as well. The Defuddle parse uses both processor time and memory (on the Java heap) roughly proportional to the input file size. The quantitative values are likely to be related to the size, type, and complexity of the input data as some of the analysis varying the row versus column dimensions of the example’s data block shows. We cannot characterize these relationships in detail from our current data.

## 5.1. Method

The Defuddle parser (prelease) was tested on amd64, 4xDual Core AMD Opteron 265 1.8GHz, 16gb RAM, Linux version 2.4.21-37.ELsmp. The code was compiled and run with Java HotSpot 64-bit Server VM, version 1.5.0_07-b03, with several settings for the run time heap size. The code was profiled with YourKit Java Profiler 7.5.11 (http://www.yourkit.com/).

The input files were variations of SimpleCSV is described above. The size of the input was varied by adding rows of data to the file to make the total size on disk to the reported size. Each run used the same parser and the same schema, differing only in the input data file. We take the raw size of the input file (bytes) as the input data size and did not distinguish the header versus data sizes.

Each run was a similar file with the same preamble and similar data, differing only by the total number of rows to read. Note that this means that, though each element can be read separately, the size of the ‘data block’ in the XML is proportional to the total size of the file.

## 5.2. Results

For small files (10MB or less), Defuddle completes in a few minutes, scaling linearly with the size of the input file. Table 1 shows the heap memory, CPU and Total time for small files. Profiling indicates that the execution time is in the parsing code, as would be expected. In particular, the time is predominately spent in the ‘marshalling’, which is generating the XML output from the data in memory. Defuddle does not seem to be I/O bound for either input or output. It appears to be memory-bound, as discussed below.

[3] The performance analysis uncovered some inefficient code that is neither essential nor difficult to replace, such as an unnecessary copy of the entire output file. This is omitted from the performance numbers here.

For files larger that about 20MB, the parser runs slower and encounters an “out of memory” exception in which the Java heap is exhausted. The behavior is consistent with thrashing: the parser requires a large amount of heap memory, forces repeated garbage collection, and eventually exhausts the heap.

The precise behavior depends on the Java virtual machine, the heap size, and garbage collection. The general trend appears to be linear for small files (< 20MB), requiring at least 50 times the file size for heap memory. That is, for each megabyte of data to parse, Defuddle uses 50MB of Java heap. While further investigation is needed, it appears that much of this memory requirement is due to the design of JaxMe itself. For example, JaxMe constructs a full in-memory data model which can then be exported as an XML file versus directly generating an XML stream. Analysis of JaxMe and of the extensions required to support DFDL could potentially uncover ways to reduce memory requirements related to the number of data copies in memory, but the linear scaling is fundamental to the JaxMe model of creating a full in-memory representation.

**Table 1. Simple CSV File Internal Parse Thread Profiling (JaxMe/jaxb), heap size 1GB.**

| Data File Size | Allocated Memory | CPU Time (hh:mm:ss) | Total Time (hh:mm:ss) |
|---|---|---|---|
| 1M | 451M | 00:02:08 | 00:02:11 |
| 2M | 342M | 00:03:08 | 00:03:12 |
| 5M | 578M | 00:08:17 | 00:08:25 |
| 10M | 856M | 00:16:12 | 00:16:27 |
| 20M | 1022M | 00:32:12 | 00:43:15 |
| 25M | 1023M | | Fail after 2 hours |
| 30M | 1023M | | Fail after 2 hours |

To investigate larger files, the Defuddle parser was run using Java64 with the heap size set to 10GB. Table 2 shows the memory used in this case The smaller input files run in similar time, though they use much more memory on the heap for the same input file (compare Table 1 and Table 2). This simple reflects the high water marks of garbage collection—in the larger heap, the garbage collection lets more memory accumulate before cleaning it up. (Neither number precisely reflects the instantaneous working set of the program.) The larger heap allows 20 and 25MB files to complete, scaling linearly with the size of the input file. The 30 MB file runs a long time and ultimately fails with “out of heap memory”.

**Table 2 Simple CSV File Internal Parse Thread Profiling (JaxMe/jaxb) vm options –Xmx10g**

| Data File Size | Allocated Memory | CPU Time (hh:mm:ss) | Total Time (hh:mm:ss) |
|---|---|---|---|
| 1M | 1010M | 00:02:04 | 00:02:13 |
| 2M | 1.9G | 00:03:08 | 00:03:47 |
| 5M | 3.4G | 00:08:04 | 00:10:00 |
| 10M | 3.6G | 00:16:11 | 00:20:07 |
| 20M | 4G | 00:32:07 | 00:33:05 |
| 25M | 4.1G | 00:40:49 | 01:41:19 |
| 30M | | Fail | |

To gain more insight into the cause of the memory exhaustion, the Defuddle parser was instrumented to report each time an integer or float is read from the text input. These traces give a direct measure of the amount of data parsed up to the point of failure. Defuddle was executed using the SimpleCSV schema and variations, with 32, 64, 128, and 256 MB of heap memory. The input file, similar to the simplecsv.txt described above, was extended to have 741,000 rows.

The results were very consistent: each run produced exactly the same result for each condition. Furthermore, for a given schema, the number of integers read until failure was linearly correlated with the heap size ($R^2 = 1.0$)[4].

However, *with the same input file*, alternative schemas produced considerably different memory usage. Specifically, the number of *rows* in the output was as important as the number of individual integers. Essentially, the schema defines the structure of the output data, which is the critical faction in memory use as the parser run.

Table 3 shows results for three schemas which differ only in the number of integers per row. The SimpleCSV schema parses above defines 10 integers per row (Figure 9 above, definition of element "item"). The middle column in Table 3 shows the results of using this schema to parse the simplecsv.txt with 741,000 rows. The heap was exhausted in each case (e.g., with the heap set at 32MB the parser can read (exactly) 454,890 integers before running out of heap). As the size of the heap is increased, the number of integers read is linear (middle line in Figure 14).

**Table 3. Data read until heap is exhausted for several heap sizes. (*This condition completed.)**

| Heap size (MB) | Items per Row | | |
|---|---|---|---|
| | 2 | 10 | 100 |
| 32 | 112,138 | 454,890 | 1,246,500 |
| 64 | 225,144 | 923,350 | 2,506,400 |
| 128 | 450,900 | 1,854,810 | 5,023,600 |
| 256 | 893,606 | 3,711,218 | 7,410,000* |

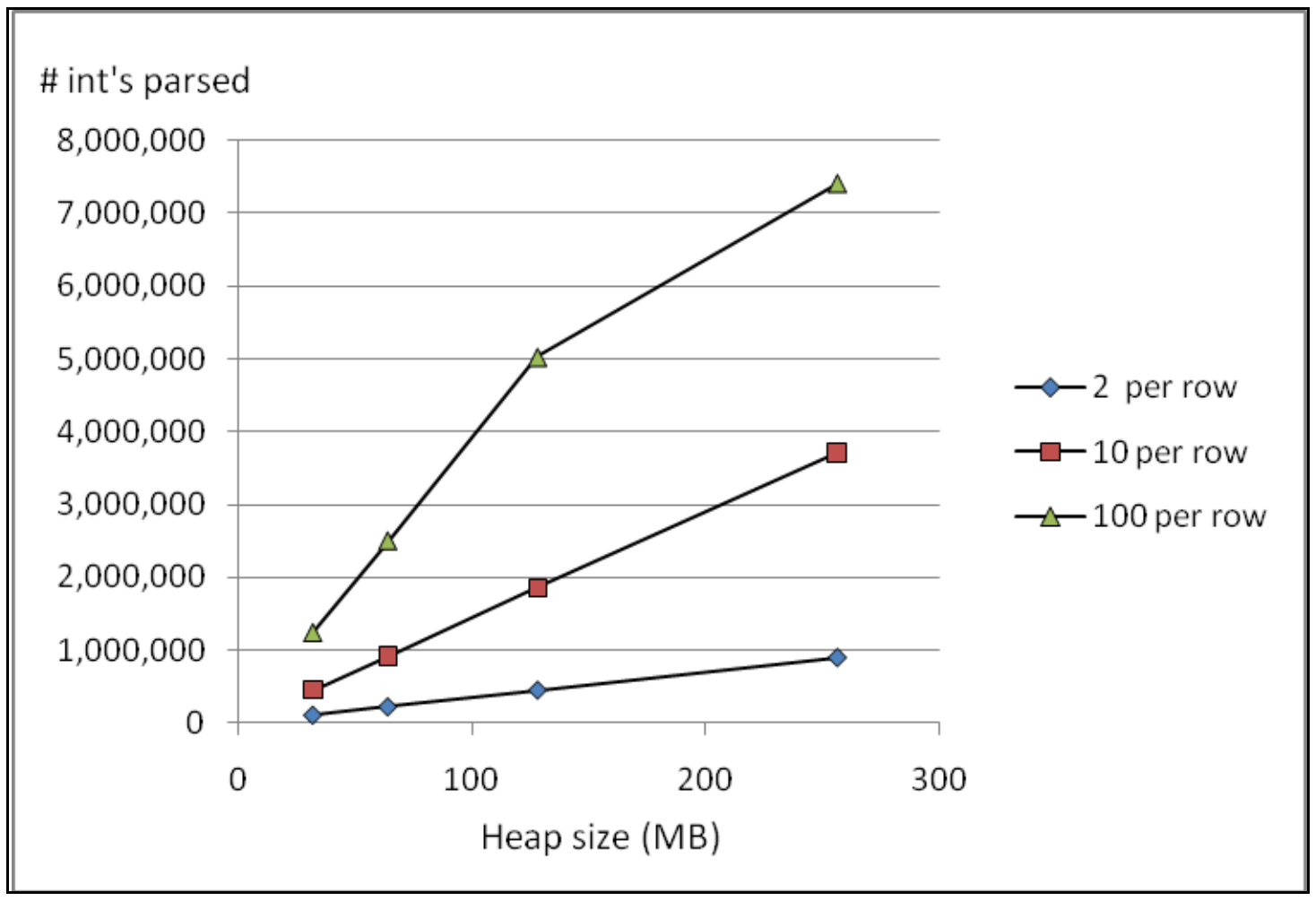

**Figure 14. Data items parsed at the point where the heap is exhausted. The input data is identical in each case, but schemas define different numbers of integers per row in the output.**

[4] The results are identical—precisely identical—for floats.

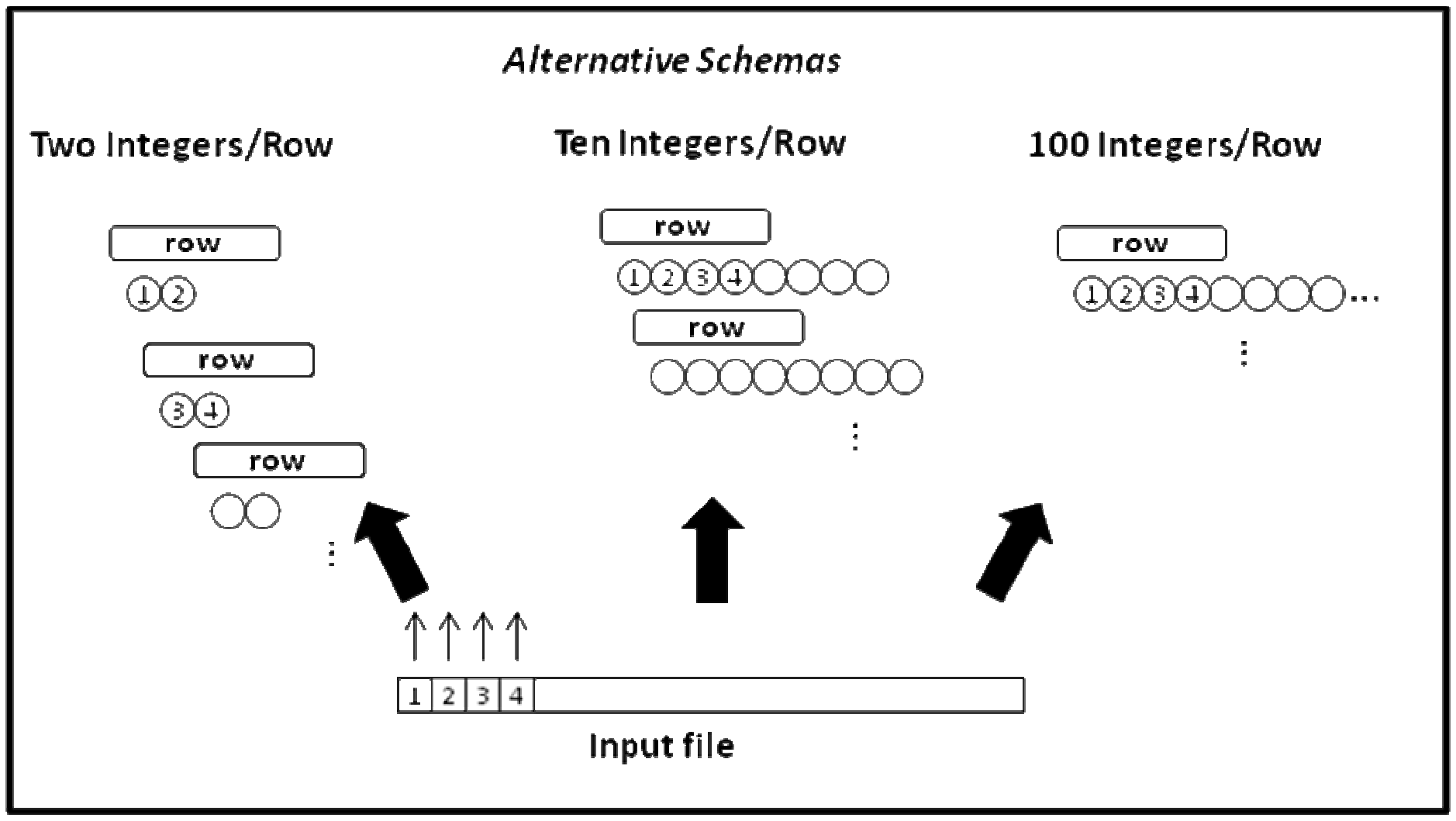

**Figure 15. Illustration of the storage required for alternative schemas.**

In comparison, a similar schema that writes 100 items per row reads far more integers from the file before exhausting the same amount of memory (right column in Table 3). In fact, this schema can parse the entire file when the heap is 256 MB. The input data is identical in each case: the difference is the schema that defines the structure of the output data. In third condition, a similar schema with only two integers per row reads far less data before exhausting the heap (left column in Table 3). Figure 14 shows the number of integers read in each condition. In each case, the trend is linear, though the slope is different.

A simple explanation for these results is that each integer from the file is read into a Java Integer object on the heap, and each row defined in the schema is stored as a Java object on the heap, which includes an ArrayList of the Integer objects. Figure 15 sketches the three cases: the input data is read into a list of lists, with different numbers of elements per list. The more integers per row, the fewer row objects to be stored. The balance of integers per row determines the total number of used on the heap, and the overhead for each row is, evidently, quite significant.[5]

## 5.3. Discussion

These results show that the Defuddle parser performs adequately for small and simple cases. The current implementation is limited by the Java heap/garbage collection during the parsing phase. The data suggest that the Defuddle parser is not I/O bound, but is memory bound.

These performance issues are due to a combination of limitations in the Java Virtual Machine (specifically, memory and garbage collection), the JaxMe parser, and the Java XML package. The reported memory usages reflect the behavior of the garbage collection in response to the demands of the parser. For the 1-10MB the memory used was far greater when the heap was 10GB than 1GB, though everything else was the same, and the run time was equivalent. This strongly suggests that the parser demands large amounts of memory as it runs, and the garbage collection is reclaiming memory for the smaller files.

The analysis of the memory usage clearly indicated that the schema determines the amount of memory used on the stack. That is, the complexity and nesting of *the desired output* determine the memory usage per record, rather than the raw number of bytes in the input file. For this reason, it is not possible to predict the heap size from the data file itself. Note that the relationship between the schema and the heap usage is complex and depends on internal details of the parser and VM (e.g., what structure is created in memory to represent a compound element).

[5] There are other overheads, but these two are completely determined by the schema.

It should be noted that the current implementation of JaxMe and Defuddle use Java JDK 1.5. The Java XML package was completely reimplemented in JDK 1.6 and later, so it is possible (though not certain) that porting to the current version of the JVM would produce substantially better performance. This update would be a logical next step in pursuing performance enhancement.

It is also important to note that the parsing was proportional to the size of input in this example only because the entire input file was extracted into the output XML. In some cases, the output may be a tiny fraction of the input. For example, the Defuddle schema might extract the header information while ignoring the data. In this case, the speed and memory use would be low and almost identical (proportional to the header size and to the output XML size) no matter how large the input file.

# 6. Implications and Future Work

The current project has updated the Defuddle parser to the current version of JaxMe and has enhanced the build and test environment. The current parser has all the features of the earlier releases, plus initial support for semantic extensions using GRDDL. Defuddle+GRDDL provides a declarative specification for extracting both data structures and semantic relations from arbitrary text and binary data, producing standard XML and RDF. An initial iRODS microservice has been developed that paves the way for integration of Defuddle into the SHAMAN preservation system. Together, these efforts have updated and extended a potent tool for preservation and have moved DFDL/Defuddle significantly closer to practical use.

As discussed above, the prototype Defuddle microservice discussed above is being developed as method to integrate Defuddle into an end-to-end demonstration is being developed by the University of Liverpool and DICE. Defuddle provides a technology by which other forms of data can be brought into this system by extracting XML and RDF representations of the non-text objects. There are a great variety of such data, with semantics at various levels of detail and specificity, Defuddle can apply alternative transformations from different logical perspectives, generating RDF which can be used with other RDF metadata. Thus, additional information generated through these uses of Defuddle can be captured for long-term use. However, there could be additional value in coupling Defuddle+GRRDL with a database-style semantic store, i.e. one that supports querying (i.e. using SPARQL [13]), inference, rule-based processing, etc. This would allow the direct outputs from Defuddle to be further processed in the overall context of collections, projects, and community knowledge bases to generate additional metadata.

Within an overall preservation environment, semantically extended Defuddle may play several unique roles. While the initial SHAMAN demonstration focuses on text documents (e.g., PDF, .doc, open document, etc.), for which there are well-understood common structural models that can be used in discovery and presentation, Defuddle provides a technology by which other forms of data can be brought into this system by extracting XML and RDF representations of the non-text objects.

The semantic extensions to Defuddle also provide a general purpose mechanism for extracting metadata about relations within the data and between multiple data objects. In the archiving system this is important for preservation of logical relationships, and for generating annotations to be used for discovery and access. For example, Defuddle might to be used to add metadata to record that data values represent "speed" in units of "meters per second" and for the GRRDL stage to create RDF that further connects these metadata to ontologies covering those topics. Note that this adds relevant semantics to the logical descriptions that are not directly retrieved from the file instance.

In addition to understanding the semantics of specific objects and their connection with reference knowledge, it is also important to capture semantics that can only be derived from the relationship of the data within the file to other files and the larger context of work. Simple examples of this could include understanding that one can potentially infer that data was created by a given author (Dublin Core "creator") from the fact that the file is associated with a "project" created by that person or is the outcome of a process run by that person.

NCSA has begun exploring uses of Defuddle to interpret files as part of other projects. These efforts include initial work to extract metadata from 3-D file formats in concert with Peter Bajcsy's effort within the same "Innovative Systems and Software: Applications to NARA Research Problems" project and work to extract common information from multiple related ASCII formats in the area of infectious disease monitoring (in collaboration with Ian Brooks at NCSA). The preliminary efforts they represent a move to a phase of assessing Defuddle in real-world use cases. In addition to the iRODS microservice, preliminary work is developing a mechanism (as a Tupelo Context) to invoke Defuddle as part of a file upload operation in Tupelo and to store the output as metadata or new content linked to the original data via provenance information (see, Yigang Zhou, Google Summer Of Code Student Fellowship, http://socghop.appspot.com/student_project/show/google/gsoc2009/ncsa/t124022775909).

With these efforts to incorporate Defuddle into NCSA's Tupelo semantic content management middleware, coupled with Tupelo-aware tools such as scientific workflow engines, provenance management interfaces, document libraries, and higher-level "digital watershed" services and web interfaces (see http://cet.ncsa.uiuc.edu/)), we can begin to see the general value beyond curation and preservation. Together with SHAMAN and iRODS, these technologies could incorporate many sources of information from ongoing scientific and engineering research and education projects into an excellent testbed for exploring the potential of rich metadata.

## 7. Acknowledgements

This work was supported through National Science Foundation Cooperative Agreement NSF OCI 05-25308 and Cooperative Support Agreements NSF OCI 04-38712 and NSF OCI 05-04064by the National Archives and Records Administration.

## 8. References

1. Brickley, Dan and Libby Miller, *FOAF Vocabulary Specification 0.9*. FOAF, 2007. http://xmlns.com/foaf/spec/
2. Connolly, Dan, *Gleaning Resource Descriptions from Dialects of Languages (GRDDL)*. W3C W3C Recommendation, 2007. http://www.w3.org/TR/grddl/

3. Dan Connolly (ed.), *Gleaning Resource Descriptions from Dialects of Languages (GRDDL)*. W3C W3C Recommendation, 2007. http://www.w3.org/TR/grddl/
4. ESW Wiki. *Grddl Implementations*. 2008, http://esw.w3.org/topic/GrddlImplementations.
5. Futrelle, Joe. *Harvesting RDF Triples*. In *IPAW'06 International Provenance and Annotation Workshop* 2006, 64-72.
6. Futrelle, Joe, Jeff Gaynor, Joel Plutchak, Peter Bajcsy, Jason Kastner, Kailash Kotwani, Jong Sung Lee, Luigi Marini, Rob Kooper, Robert McGrath, Terry McLaren, Yong Liu, and James Myers, *Knowledge Spaces and Scientific Data (Poster)*, in *4th IEEE International Conference on e-Science*. 2008: Indianapolis.
7. Futrelle, Joe and James Myers, *Tracking Provenance in Heterogeneous Execution Contexts.* Concurrency and Computation: Practice and Experience, *20 (5):555-564*, 10 April 2008.
8. Lagoze, Carl, Herbert Van de Sompel, Pete Johnston, Michael Nelson, Robert Sanderson, and Simeon Warner, *ORE User Guide - Resource Map Implementation in RDF/XML*. Open Archives Initiative, 2008. http://www.openarchives.org/ore/rdfxml
9. Moore, Regan, *email*. 2008.
10. Moore, Regan and MacKenzie Smith. *Assessment of RLG Trusted Digital Repository Requirements*. In *Workshop on "Digital Curation & Trusted Repositories: Seeking Success" at Joint Conference on Digital Libraries (JCDL 2006)*, 2006. http://sils.unc.edu/events/2006jcdl/digitalcuration/
11. Moreau, Luc, Juliana Freire, Joe Futrelle, Robert McGrath, Jim Myers, and Patrick Paulson, *The Open Provenance Model: An Overview*, in *Provenance and Annotation of Data and Processes*. Springer, Berlin, 2008, 323-326. http://dx.doi.org/10.1007/978-3-540-89965-5_31
12. Ogbuji, Chimezie, *GRDDL Test Cases*. W3C Recommendation, 2007. http://www.w3.org/TR/grddl-tests/
13. Prud'hommeaux, Eric and Andy Seaborne, *SPARQL Query Language for RDF*. W3C W3C Recommendation, 2008. http://www.w3.org/TR/rdf-sparql-query/
14. SHAMAN. *Sustaining Heritage Access through Multivalent ArchiviNg*. 2008, http://www.shaman-ip.eu/.
15. Talbott, Tara D., Karen L. Schuchardt, Eric G. Stephan, and James D. Myers, *Mapping Physical Formats to Logical Models to Extract Data and metadata: The Defuddle Parsing Engine*, in *International Provenance and Annotation Workshop*. 2006, Springer: Heidelberg. p. 73-81.
16. W3C. *GRDDL Working Group*. 2008, http://www.w3.org/2001/sw/grddl-wg/.
17. Watry, Paul, *Digital Preservation Theory and Application: Transcontinental Persistent Archives Testbed Activity*. The International Journal of Digital Curation, *2 (2):41-68*, November 2007. http://www.ijdc.net/ijdc/article/view/43
18. Watry, Paul, *email*. 2008.

# Appendix A. The Full Schema for Example

```
<?xml version="1.0" encoding="UTF-8"?>
<xs:schema elementFormDefault="qualified"
attributeFormDefault="unqualified"
xmlns:xs="http://www.w3.org/2001/XMLSchema"
targetNamespace="Dataset" xmlns="Dataset" xmlns:dataset="Dataset"
xmlns:dfdl="DFDL">

    <xs:element name="table" type="SimpleTable">
    </xs:element>

    <xs:complexType name ="SimpleTable">
        <xs:sequence>
            <xs:element name="hdrblock" type="header" />
            <xs:element name="datablock" type="Row"
maxOccurs="unbounded"/>
        </xs:sequence>
    </xs:complexType>

    <xs:complexType name="header">
        <xs:sequence>
            <xs:annotation>
                <xs:appinfo>
                    <dfdl:dataFormat>
                        <dfdl:repType>text</dfdl:repType>
                    </dfdl:dataFormat>
                </xs:appinfo>
            </xs:annotation>
            <xs:element name="Author" type="xs:string" >
                <xs:annotation>
                    <xs:appinfo>
                        <dfdl:dataFormat>
            <dfdl:ignore>Creator:\\s</dfdl:ignore>
                            <dfdl:terminator
           kind="regexp or string">\\r\\n|[\\r\\n]</dfdl:terminator>
                        <dfdl:charset>US-ASCII</dfdl:charset>
                        </dfdl:dataFormat>
                    </xs:appinfo>
                </xs:annotation>
            </xs:element>
            <xs:element name="CreationDateDate" type="xs:string" >
                <xs:annotation>
                    <xs:appinfo>
                        <dfdl:dataFormat >
            <dfdl:ignore>Date:\\s</dfdl:ignore>
                            <dfdl:terminator
              kind="regexp or
string">\\r\\n|[\\r\\n]</dfdl:terminator>
                        <dfdl:charset>US-ASCII</dfdl:charset>
                        </dfdl:dataFormat>
                    </xs:appinfo>
                </xs:annotation>
            </xs:element>
        </xs:sequence>
```

```
    </xs:complexType>

    <xs:complexType name="Row">
        <xs:sequence>
            <xs:annotation>
                <xs:appinfo>
                    <dfdl:dataFormat>
                        <dfdl:repType>text</dfdl:repType>
                    </dfdl:dataFormat>
                </xs:appinfo>
            </xs:annotation>
            <xs:element name="item" type="xs:int" minOccurs="10"
maxOccurs="10">
                <xs:annotation>
                    <xs:appinfo>
                        <dfdl:dataFormat xmlns:dfdl="DFDL">
                            <dfdl:separator
     kind="regexp or string">,\\r\\n|,[\\r\\n]|,|\\r\\n|[\\r\\n]
                                         </dfdl:separator>
                            <dfdl:base>10</dfdl:base>
                        <dfdl:charset>US-ASCII</dfdl:charset>
                        </dfdl:dataFormat>
                    </xs:appinfo>
                </xs:annotation>
            </xs:element>
        </xs:sequence>
    </xs:complexType>
    <xs:complexType name="Creat">
        <xs:sequence>
            <xs:annotation>
                <xs:appinfo>
                    <dfdl:dataFormat>
                        <dfdl:repType>text</dfdl:repType>
                    </dfdl:dataFormat>
                </xs:appinfo>
            </xs:annotation>
            <xs:element name="c" type="xs:string">
                <xs:annotation>
                    <xs:appinfo>
                        <dfdl:dataFormat xmlns:dfdl="DFDL">
                            <dfdl:terminator
                          kind="regexp or
string">:\\s</dfdl:terminator>
                            <dfdl:separator
                           kind="regexp or
string">:\\s</dfdl:separator>
                        <dfdl:charset>US-ASCII</dfdl:charset>
                        </dfdl:dataFormat>
                    </xs:appinfo>
                </xs:annotation>
            </xs:element>
        </xs:sequence>
    </xs:complexType>
</xs:schema>
```